\documentstyle[prl,aps,epsf]{revtex}

\begin{document}
\bibliographystyle{prsty}
\twocolumn[\hsize\textwidth\columnwidth\hsize\csname@twocolumnfalse%
\endcsname
\draft
\preprint{SU-ITP \# 97/33}
\title{Dispersion of the $\pi$ resonance}
\author{Jiang-Ping Hu and Shou-Cheng Zhang}

\address{\it Department of Physics, McCullough Building, Stanford
  University, Stanford  CA~~94305-4045 \\
and\\
 Institute for Theoretical Physics, University of California, Santa Barbara, CA 93106}
\maketitle
\begin{abstract}
\end{abstract}
{\sl We study the dispersion of the $\pi$ resonance in the superconducting
state within the projected $SO(5)$ model\cite{project}. Away from the
the commensurate momentum, the propagation of the $\pi$ resonance creates
phase slips in the superconducting order parameter. This frustration effect
leads to a strong dressing of the $\pi$ resonance and a downwards dispersion
away from the commensurate wave vector. Based on these results, we argue that the
the commensurate resonance and incommensurate magnetic fluctuations in the cuprates
are continuously connected.}

 ]

The emergence of a commensurate neutron resonance peak below $T_c$
is one of the most striking properties of the cuprates. There are
many different theoretical approaches to the problem. Within the
$SO(5)$ theory\cite{so5}, this resonance peak is interpreted as a collective
mode describing a rotation from the superconducting (SC) state to
the commensurate antiferromagnetic (AF) state. Since the rotation
operator is a triplet particle-particle operator, which can only
couple to neutrons in a SC state, this theory predicts that the
intensity of the $\pi$ resonance mode is proportional to the
square of the SC order parameter\cite{neutron}. This fact naturally explains the
temperature dependence of the resonance intensity observed in
experiments\cite{keimer}, and anticipated\cite{zeeman} the recently observed dependence of
the resonance intensity on the magnetic field\cite{field}. A c-axis magnetic
field creates non-SC vortex cores, therefore it reduces the SC
order parameter on the average. On the other hand, a magnetic
field aligned in the ab plane has little effects in reducing the
SC order parameter. From the dependence of the resonance intensity
on the SC order parameter within the $SO(5)$ theory, one would
therefore naturally expect a reduction of the resonance intensity
in the presence of a $c$ axis magnetic field.

More recently, incommensurate magnetic fluctuations have also been
observed in the YBCO superconductors\cite{dai,keimer2}. Unlike their counterparts in the
LSCO superconductors, these incommensurate magnetic fluctuations
are energy dependent, and seem to merge continuously into the $\pi$
resonance mode. In fact, this behavior is predicted by Furusaki and
one of us\cite{furusaki}, who applied the $SO(5)$ theory to the ladder systems.
Using a asymptotically exact renormalization group method, they
found that the $\pi$ resonance mode in the ladder system has a downward
dispersion away from the commensurate wave vector (note the
remarkable resemblance of Fig. 5 and 6 of
reference \cite{furusaki} and the experimental Fig. 3A of Ref. \cite{keimer2}).
In fact, in a earlier work, Poilblanc, Scalapino and
Hanke\cite{poilblanc} found similar behavior in a exact diagonalization study of
the ladder systems. The purpose of this work is to identify the
physical origin of this downwards dispersion relation and construct
a unified theory of the commensurate and incommensurate magnetic
fluctuations in the cuprates.

To understand the basic physics of the competition between the
commensurate and incommensurate magnetic fluctuations, let us
first consider a very simple system of doped $t-J_{||}-J_{\perp}$
ladder. In the $J_{\perp}>>J_{||}$ limit,
the SC state of the doped ladder system can be viewed as a
coherent linear superposition of a hole pair and a singlet pair on
each rung. In this picture, the $\pi$ resonance is simply a linear
superposition of a spin triplet excitation on each rung. Because
the SC state is a broken symmetry state, such a excitation can be
either created by a spin triplet operator or a triplet
particle-particle operator, the so called $\pi$ operator. Simple
second order perturbation theory shows that the exchange parameter
between the triplet and a hole pair is
$-\frac{4t^2}{J_{\perp}}$, which has a negative sign, while the
exchange parameter between the triplet and the singlet pair is
$\frac{J_{||}}{2} $, which has a  positive sign. The positive
exchange sign prefers a minimal energy wave vector at
$q_{min}=\pi$, while the negative sign prefers a minimal
energy wave vector at $q_{min}=0$. Since the SC ground state
is a linear superposition of both the hole pair and the singlet pair,
this effect leads to a competition between these two band minima.
In one dimension, this competition always leads to a
incommensurate band minimum. This picture is a crude caricature of
the basic physics leading to the incommensurate band minimum for
the $\pi$ resonance. For the full calculations in one dimensions,
readers are referred to the original paper by Furusaki and one of
us\cite{furusaki}.

In this paper, we shall generalize this physics to two dimensions
and study the propagation of the $\pi$ resonance in the SC state,
within the projected $SO(5)$ model ($pSO(5)$ model). (In fact, within the
pure SC state, there is no essential difference between the $SO(5)$ model
and the projected $SO(5)$ model\cite{project}.) Just like the physical picture
outlined above, the propagation of the $\pi$ resonance leads to a
sign reversal of the SC order parameter behind it. Therefore, the
problem is similar to the problem of the propagation of a single
hole in the antiferromagnetic background, which has been studied
extensively in the literature\cite{siggia,trugman,kane,liu}.
In fact, using basically the same approximations, we shall show that
the quantum fluctuations erase the string of sign reversals, but
leads to a non-trivial correction to the dispersion of the
$\pi$ resonance. Depending on parameters of the model, the
correction can give rise to a downward dispersion of the $\pi$
mode, reaching a minimum at incommensurate wave vector.
Current experiments only show
a downwards dispersion of the $\pi$ resonance. The main prediction
of our theory is that there will be a upward turn in the dispersion
after the minimum is reached.

\begin{figure*}[h]
\centerline{\epsfysize=3.0cm \epsfbox{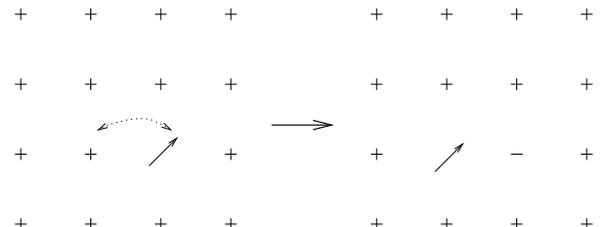} } \caption{
Schematic diagram of the hopping of a magnon in SC state. The arrow
represents a magnon while the the signs represent the phase of SC order. }
\label{fig1}
\end{figure*}

We begin with the projected  $SO(5)$ model defined on a lattice,
(using the notations of Ref. \cite{project}.)
\begin{eqnarray}
H &=&
\Delta_s \sum_{ x } t_\alpha^\dagger(x) t_\alpha(x) +
\tilde \Delta_c \sum_{ x } n_i(x) n_i(x)  \nonumber \\
&-& J_s \sum_{ <xx'> } n_\alpha(x) n_\alpha(x') -
J_c \sum_{ <xx'> } n_i(x) n_i(x') \nonumber \\
&+& V\sum_{<xx'>}L_{ab}(x)L_{ab}(x') \label{hamiltonian}
\end{eqnarray}
where $i=1,5$, $\alpha = 2,3,4$ and $\tilde \Delta_c = \Delta_c -\mu$ where
$\mu$ is the chemical potential.  In $SO(5)$ superspin notation, the superspin
is defined as
\begin{eqnarray}
n_1 &=& \frac{1}{2} (t_h + t_h^\dagger) \ \ \
n_5 = \frac{1}{2i} (t_h - t_h^\dagger) \\
n_{\alpha} &=& \frac{1}{\sqrt{2}}(t_{\alpha} +t^+_{\alpha}).
\label{after}
\end{eqnarray}
where $t_h$ and $t_{\alpha}$ are hard-core boson annihilation
operators for the hole pair and the magnon. Here one lattice site
of this effective model corresponds to a plaquette in the original
$C_uO_2$ plane, and we have made a shift of the momentum vector
for $t_\alpha$ bosons by $(\pi,\pi)$. $pSO(5)$ model
\cite{project} is constructed by projecting out doubly occupied
configuration from the local $SO(5)$ multiplets. Even though some
members of the $SO(5)$ multiplets are projected out, at
$\Delta_s=\tilde\Delta_c$ and $J_c=2J_s$, the mean field ground
state manifold is still $SO(5)$ symmetric, and AF can be smoothly
rotated into SC with no energy cost. In Ref.\cite{project}, we
have shown this model gives a realistic description of the global
phase diagram of the cuprates and many of their
physical properties. In that paper, the $V$-term in above
Hamiltonian was simply ignored because it does not impact the mean field
phase diagram if only pure AF phase and SC phase are concerned.
However, one important observation in this paper is that this term
plays important role on the dispersion of $\pi$ mode. If we limit
our discussion on one single magnon in Hilbert space, the $V$-term  simply
describes the hopping exchange between a hole-pair and a magnon.
The hopping between a magnon and a singlet is described in the
term $-J_sn_{\alpha}(x)n_{\alpha}(x')$. Since we only discuss the
motion of an single magnon, all of other matrix elements for these
two terms vanish  except
\begin{eqnarray}
V L_{ab}(x)L_{ab}(x')(t^+_h(x)t^+_\alpha(x')|0>) &=& V t^+_h(x')t^+_\alpha(x)|0> \nonumber \\
-J_s n_{\alpha}(x)n_{\alpha}(x')(b^+(x)t^+_\alpha(x')|0>) &=& -J_s b^+(x')t^+_\alpha(x)|0>,
\end{eqnarray}
where $b^+(x)$ creates an singlet at site $x$. Since both of $J_s$ and $V$ is
 positive in this model, we see that these two hopping processes have opposite signs.
Therefore, in a pure SC state, which is a coherent state
 of singlet and hole pair on each site, the hopping of magnon flips the phase of the
 local SC order. More precisely, let's take a mean field pure SC
 state, $\Phi$,
\begin{eqnarray}
\Phi = \prod_{x}\phi_{+}(x) , ~~\phi_{\pm}(x)=(cos\theta b^+(x) \pm sin\theta t_h^+(x))|0>,
\label{wave}
\end{eqnarray}
which has SC order, $<n_1(x)> = \frac{1}{2}sin(2\theta)$, on
each site. Taking the hopping term of a magnon between sites $x,x'$
we obtain for $V=J_s$,
\begin{eqnarray}
(L_{ab}(x)L_{ab}(x')-n_{\alpha}(x)n_{\alpha}(x')) t^+_{\alpha}(x')\phi_{+}(x)
= t^+_{\alpha}(x)\phi_{-}(x'),
\end{eqnarray}
in which the SC order at site $x'$,  $n_1(x')$, takes value
$<n_1(x)> = -\frac{1}{2}sin(2\theta)$. A schematic diagram is
shown in Fig.\ref{fig1} to reflect the hopping of a magnon in the SC
state. The motion of  a magnon creates the phase mismatch similar
to the spin mismatch caused by the motion of a single hole in a AF state.
In the Ising limit, the motion of magnon creates a phase mismatched
string and feels a linear potential which results in an infinite
effective magnon mass. However, similar to a single hole problem,
quantum phase fluctuations in the SC state can erase
the string effect and modify the dispersion relation.
We employ the same method in ref.\cite{kane,liu} to
solve the problem.

Taking the variational wavefunction in Eq.\ref{wave}, a simple
mean field calculation gives the hole-pair density
\begin{eqnarray}
\rho=sin^2(\theta)=\frac{J_cz -\tilde \Delta_c}{2J_cz},
\end{eqnarray}
 where $z$ is the coordination number on the lattice. We introduce a
Lagrangian multiplier field $\lambda$ to enforce the hard-core
boson constraint on average. To obtain the collective phase mode,
we define two new  boson operators,
\begin{eqnarray}
 \alpha_1(x) = sin\theta b(x)- cos\theta t_h(x) , \nonumber \\
  \alpha_2(x) = cos\theta b(x)+ sin\theta t_h(x),
\end{eqnarray}
 which  satisfy  $\alpha_1(x)|\Phi> = 0, <\alpha_2(x)> = 1$.
The quadratic Hamiltonian describing the phase fluctuation above
the mean field state is therefore given by
\begin{eqnarray}
 H' &=& H - <H> \nonumber \\
  &=& -J_cz\sum_{q}[(1-\gamma(q)+\frac{sin^2(2\theta)}{2}
 \gamma(q))\alpha_1^+(q)\alpha_1(q) \nonumber \\
 &+& \frac{sin^2(2\theta)}{4} \gamma(q)(\alpha_1(-q)\alpha_1(q)+h.c.)]
\end{eqnarray}
where $\gamma(q) = \frac{cos(q_x)+cos(q_y)}{2}$. In the following
part of paper, we  normalize all of the energy scale by taking
$(J_s+V)z/2 =1$ and define two additional dimensionless
parameters,
\begin{eqnarray}
x = \frac{V-J_s}{J_s+V}, ~~ y = \frac{ 2J_c}{(J_s+V)}.
\end{eqnarray}
 By standard diagonalization,
\begin{eqnarray}
H' = \sum_{q} \epsilon(q)\beta^+(q)\beta(q),
\label{hphase}
\end{eqnarray}
where
\begin{eqnarray}
 \beta(q) &=& u(q)\alpha_1(q) - v(q)\alpha_1^+(-q) \\
\label{beta}
\epsilon(q) &=& y\sqrt{(1-\gamma(q))^2+sin^2(2\theta)\gamma(q)(1-\gamma(q))} \\
 u(q) &=& \sqrt{\frac{1}{2}+y\frac{1-\gamma(q)+\frac{sin^2(2\theta)}{2}\gamma(q))}
 {2\epsilon(q)}} \\
v(q) &=&
-sign(\gamma(q))\sqrt{-\frac{1}{2}+y\frac{1-\gamma(q)+\frac{sin^2(2\theta)}
{2}\gamma(q)}{2\epsilon(q)}}. \label{disp}
\end{eqnarray}

The hopping of a single magnon can be described by the following Hamiltonian,
\begin{eqnarray}
H_m = &-&J_s\sum_{xx'}(t_\alpha^+(x)t_\alpha(x')b^+(x')b(x)+h.c.) \nonumber \\
&+&V\sum_{xx'}(t_\alpha^+(x)t_\alpha(x')t_h^+(x')t_h(x)+h.c.).
\label{hamilt}
\end{eqnarray}
Taking the mean field expectations and using relations of
Eq.\ref{wave}, we get
\begin{eqnarray}
 H_m &=&  \sum_{q}(E_b(q,\theta)t_\alpha^+(q)t_\alpha(q)+h.c.) \nonumber \\
 &+& \sum_{kq}f(k,q)t_\alpha^+(k)t_\alpha(k-q),
 \label{hm}
\end{eqnarray}
where
\begin{eqnarray}
E_b(q,\theta)=(x-cos(2\theta))\gamma(q).
\end{eqnarray}
and
\begin{eqnarray}
f(k,q)&=& sin(2\theta)[(\gamma(k-q)u(q)+\gamma(k)v(q))\beta(q) \nonumber \\
&+&(\gamma(k)u(q)+\gamma(k-q)v(q))\beta^+(-q)].
\end{eqnarray}
Given the Hamiltonian composed of Eq.\ref{hphase} and
Eq.\ref{hm}, we can  calculate the dynamic spin correlation function,
\begin{eqnarray}
G(x,t)  =
-i<\Phi|T[t_\alpha(x,t)b^+(x,t)t^+_\alpha(0,0)b(0,0)]|\Phi>.
\end{eqnarray}
Taking the mean field value on $b$ and using the self-consistent
perturbation which sums only noncrossing diagrams, we obtain the
following Dyson's equation,
\begin{eqnarray}
    G(k,\omega) = \frac{cos^2(\theta)}{\omega-E_0(k,\theta)-\sum_qF(k,q,\theta)
            G(k-q,\omega-\epsilon(q))}
   \label{dyson}
 \end{eqnarray}
where
 \begin{eqnarray}
    F(k,q,\theta) = sin^2 (2\theta)|\gamma(k-q)u(q)+\gamma(k)v(q)|^2.
 \end{eqnarray}
In this approximation, the vertex is neglected, which is small
according to the calculation in Ref.\cite{liu} for the single hole
problem.  In our model, the problem is just more
complicated in the sense that the coupling strength and the
dispersion of phase mode are quite different and depend on the
density of hole pairs. We numerically solve the above integral
equations to obtain the spectrum, $A(k,\omega) =
-\frac{1}{\pi}G(k,\omega)$ , and in particular the minimum
position of the dispersion in the broad arrange of parameters. We
first observe is that  the bare dispersion of magnon, $E_b(k)$,
can be removed in Eq.\ref{dyson} by shifting $\omega$. Moreover,
if $y$ is large, it is better to rescale  $\epsilon(q)$ in
numerical calculation, which can narrow the
 energy arrange required to find a  solution. Combine these two
 observations, the numerical convergence is rather fast.

\begin{figure*}[h]
\centerline{\epsfysize=5.0cm \epsfbox{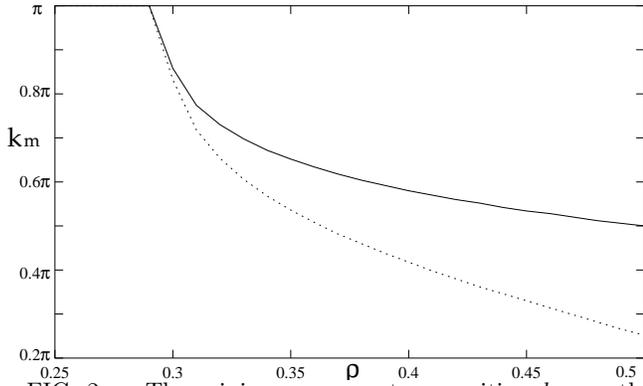} } \caption{
The minimum momentum position $k_m$ as the function of density
$\rho$ at parameters: $x=0$, $y=0.5$. Solid line reflects
$(\pi,\pi)$ direction and   dashed line represents $(\pi,0)$
direction. (we have made a shift back $(0,0)$ to $(\pi,\pi)$ in
all figures in this paper) }
 \label{fig2}
\end{figure*}

\begin{figure*}[h] \centerline{\epsfysize=5.0cm
\epsfbox{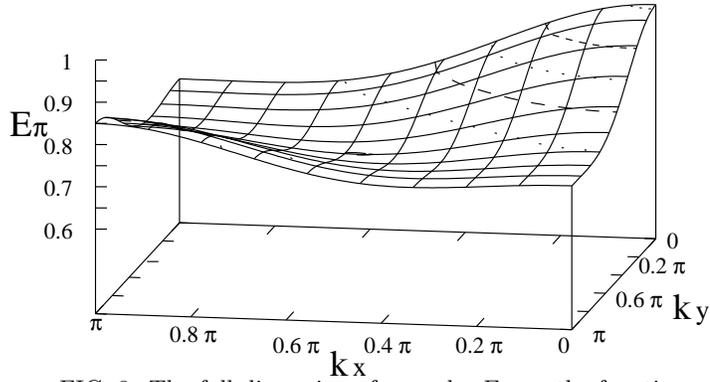} } \caption{The full dispersion  of $\pi$ mode,
$E_{\pi}$, as the function of $k_x, k_y$ at density $\rho=0.4$
with the parameters: $x=0$, $y=0.5$, $z=0.5$, $\delta_s=1$. }
\label{fig3}
\end{figure*}

Throughout the numerical calculation, a sharp coherent peak is
found in all momentum space. The spectral function can be
generally written as
\begin{eqnarray}
 A(k,\omega,\theta,y) =
cos^2(\theta)Z(k,\theta,y)\delta(\omega-\Omega(k,\theta,y))+ A',
\end{eqnarray}
where $A'$ is the incoherent part and $\Omega(k,\theta,y)$ defines
the energy dispersion.
 From our
numerical result, we find that the energy dispersion, $\Omega$, can be
written as
\begin{eqnarray}
\Omega(x,k,\theta,y) &=& \Omega_0(x,0,\theta,y)+(E_b(k,\theta)-E_b(0,\theta))\nonumber \\
&-&\delta\Omega(k,\theta,y),
 \label{omega}
\end{eqnarray}
where the first term is the energy at $k=(0,0)$, the second one is
the bare dispersion, and the third is the relative energy shift
contributed by fluctuation correction.  Based on the symmetry of
our model, $\delta\Omega(k,\rho,y) =\delta\Omega(k,1-\rho,y)$). We
find that $\delta\Omega(k,\theta)$ fits very well
a product of two separated functions, $h(\theta,y)$ and
$g(k)$, where $h(\theta,y)$ is independent of $k$.
  Considering the symmetry of
 the lattice, we find the function $g(k)$ can be well fitted to the following form,
\begin{eqnarray}
 g(k_x,k_y) &=& 0.128(sin^2(k_x)+sin^2(k_y))\nonumber \\
  &+& 0.054(cos(k_x)-cos(k_y))^2.
\label{gk}
\end{eqnarray}
 Thus, the minimum of $\Omega$ can be analytically determined by
plugging  Eq.\ref{gk} into Eq.\ref{omega}. The global minimum
always happens along the diagonal $(\pi,\pi)$ direction. However,
along the $(\pi,0)$ direction, there is a local minimum. The
global and local minimum
 positions  start to shift from $(0,0)$ at a common critical density $\rho_I$ which is
approximately determined by the equation
\begin{eqnarray}
cos(2\theta_I)-0.512h(\theta_I,y)-x=0.
\end{eqnarray}
In Fig.\ref{fig2}, we show  the global minimum position,
$(k_m,k_m)$, and local minimum, $(k_m,0)$, as the function of the
density.  In Fig.\ref{fig3}, we also show the full dispersion of
$\pi$ mode at the density $\rho=0.4$.

There is another important density in the problem.
Within the $pSO(5)$ model, uniform SC state can only
exist for densities exceeding $\rho_c$.
Ignoring small quantum corrections, $\rho_c$  is given by
\begin{eqnarray}
\rho_c = \frac{r-\delta_s+0.5(1-x)}{0.5y+r+1},
\end{eqnarray}
where $r = J_s/(J_s+V)$ and $\delta_s = \frac{\Delta_s}{J_sz+Vz}$.
Therefore, depended on the parameters, $\rho_I$ could be larger or
smaller than $\rho_c$. For example, at $x=0$, $y=0.5$ and
$r=0.25$, $\rho_c > \rho_I$ for $\delta_s<0.3$ and
$\rho_c < \rho_I$ for $\delta_s>0.3$.
Therefore, the relative sizes between $\rho_I$ and $\rho_c$ could tell us the
nature of the states as doping level is reduced.
For $\rho<\rho_c$, a mix state
between AF and SC is obtained. If $\rho_I>\rho_c$, the lowest energy magnetic
fluctuations are commensurate at the transition, and a uniform mix state
between AF and SC is obtained. On the other hand, if $\rho_I<\rho_c$,
the lowest magnetic excitations are incommensurate at the transition,
and a stripe state with alternating AF and SC order appears.

Finally we would like to compare our results with two classes of theory of
the $\pi$ resonance, one based on particle-particle picture\cite{neutron,so5}
and the other
based on the particle hole picture\cite{norman}. Since all calculations are carried out
in the SC state, where these two channels mix, distinctions can only be
made meaningfully by comparing the final physical predictions.
In this work, the intensity of the peak is determined is $cos^2(\theta)Z(k,\theta)$,
where $Z(k,\theta)$ is basically independent of the density parameter $\theta$.
Therefore, the intensity of the magnetic collective mode is
\begin{eqnarray}
cos^2(\theta)=1-\rho=\frac{|<n_1>|^2}{\rho},
\end{eqnarray}
where $<n_1>$ is the SC order parameter. This result agrees exactly with the
prediction of the particle-particle picture\cite{neutron,kohno}.
The energy of the collective mode is independent of the SC energy
gap, and is expected to be temperature independent. Both these properties
are in contrast with the predictions of based on the particle-hole
picture, where the mode intensity is insensitive to the SC order, and
the mode energy is always less than the SC pairing gap.
The particle-hole based pictures\cite{norman} can also explain the downward dispersion,
however the mode terminates at certain wave vector due to Landau damping.
On the other hand, our current work predicts a minimum in the dispersion.
The energy at the minimum is the spin gap in the system.

We would like to acknowledge useful discussions with Drs. A. Auerbach,
E. Demler, A. Dorneich and W. Hanke,
SCZ and JP Hu are supported by the NSF under grant numbers DMR-9814289.
JP Hu is also supported by the Stanford Graduate fellowship and ITP Graduate
Fellowship in UCSB. Part of the work was carried out during the workshop
on high $T_c$ superconductivity at the Institute for Theoretical Physics
at UC Santa Barbara.

\end{document}